\documentclass[preprintnumbers, prd, showpacs, floatfix,onecolumn,
preprintnumbers, letterpaper,
superscriptaddress,nofootinbib]{revtex4}
\usepackage{amsfonts}
\usepackage{amsmath}
\usepackage{amssymb,epsf}
\usepackage{latexsym}
\usepackage{graphicx,epsfig}
\usepackage{amssymb}
\usepackage{subfigure}
\usepackage[colorlinks=true,citecolor=blue,linkcolor=blue,urlcolor=black]{hyperref}

\begin{document}

\title{Thermodynamics of charged rotating dilaton black branes coupled to \\
logarithmic nonlinear electrodynamics}
\author{A. Sheykhi}
\email{asheykhi@shirazu.ac.ir}
\author{M. H. Dehghani}
\email{mhd@shirazu.ac.ir}
\affiliation{Physics Department and Biruni Observatory, College of Sciences, Shiraz
University, Shiraz 71454, Iran}
\affiliation{Research Institute for Astronomy and Astrophysics of Maragha (RIAAM), P.O.
Box 55134-441, Maragha, Iran}
\author{M. Kord Zangeneh}
\email{mkzangeneh@shirazu.ac.ir}
\affiliation{Physics Department and Biruni Observatory, College of Sciences, Shiraz
University, Shiraz 71454, Iran}

\begin{abstract}
We construct a new class of charged rotating black brane solutions in the
presence of logarithmic nonlinear electrodynamics with complete set of the
rotation parameters in arbitrary dimensions. The topology of the horizon of
these rotating black branes are flat, while, due to the presence of the
dilaton field the asymptotic behaviour of them are neither flat nor
(anti)-de Sitter [(A)dS]. We investigate the physical properties of the
solutions. The mass and angular momentum of the spacetime are obtained by
using the counterterm method inspired by AdS/CFT correspondence. We derive
temperature, electric potential and entropy associated with the horizon and
check the validity of the first law of thermodynamics on the black brane
horizon. We study thermal stability of the solutions in both canonical and
grand canonical ensemble and disclose the effects of the rotation parameter,
nonlinearity of electrodynamics and dilaton field on the thermal stability
conditions. We find the solutions are thermally stable for $\alpha<1$, while
for $\alpha>1$ the solutions may encounter an unstable phase where $\alpha$
is dilaton-electromagnetic coupling constant.
\end{abstract}

\maketitle

\section{Introduction}

Thermodynamics of black holes plays a central role in the attractive modern
method relating classical gravity and high energy physics namely
gauge/gravity duality. The issue was first taken under consideration by
Bekenstein \cite{Bek} and Hawking \cite{Haw} and encountered increasing
interests rapidly. Among different frameworks, the thermodynamics of black
solutions has been studied within, dilaton gravity possesses a significant
predominancy. This preference has at least two reasons. From one side, the
dilaton gravity which is one of the modified gravities, is able to justify
the accelerating expansion of the Universe confirmed from observations \cite%
{ExpUni} {while Einstein gravity (General Relativity)} {%
requires exotic matter violating energy conditions to justify this phase of
universe.} From another side, the dilaton gravity appears in the low energy
limit of string theory \cite{Green} and therefore can provide a good
laboratory for testing this theory in the low energy limit through
gauge/gravity duality. Since string theory proposes higher than four
dimensions \cite{Green}, it is natural to consider higher-dimensional
solutions within the gravity theories come from string theory.

Exact asymptotically flat solutions of Einstein-Maxwell-dilaton gravity have
been constructed in the absence of dilaton potential in \cite%
{Gibbons,Gar,Mig,Pol}. However, breaking of spacetime supersymmetry in ten
dimensions may cause one or more Liouville-type potentials in the action of
dilaton gravity. This type of dilaton potential change the asymptotic
behavior of solutions \cite{MW,Cai,yaz2,CHM,Clem}. In general, these
solutions are neither asymptotically flat nor (anti) de Sitter [(A)dS].
Thermodynamics of topological dilaton black holes in Einstein-Maxwell
gravity has been explored in \cite{Shey3}. Asymptotically non-flat and
non-(A)dS linearly charged rotating black branes was taken under
investigation in \cite{rotBB}. Slowly rotating charged black holes have been
studied from thermodynamics point of view as well \cite{Slrot}.

A natural interesting extension of such solutions is to change the
electrodynamics Lagrangian from linear Maxwell to nonlinear ones. Some
efforts have been done to construct exact solution in dilaton gravity with
nonlinear electrodynamics. For example, thermodynamics of static black hole
solutions in the presence of nonlinear power-law Maxwell (PLM) \cite{topPLM}%
, Born-Infeld (BI) \cite{topBI} and exponential \cite{HendiJHEP,BHExp}
electrodynamics have been investigated. As well, thermodynamic properties of
rotating black brane solutions have been studied in the presence of PLM \cite%
{rotPLM}, BI \cite{rotBI} and exponential \cite{rotExp} nonlinear
electrodynamics. Any one of the above mentioned nonlinear electrodynamics
has its own importance and motivations. For instance, the PLM
electrodynamics extends the conformal invariance property of Linear Maxwell
Lagrangian in four dimensions to higher dimensional spacetimes, while BI
\cite{BI} electrodynamics, which comes from open string theory \cite{BI1,BI2}%
, solves the problem of infinite self-energy of charged point-particle that
appears in linear Maxwell case. The latter problem is also overcome by
logarithmic nonlinear electrodynamics proposed for the first time in \cite%
{Soleng}. Although this type of nonlinear electrodynamics has no direct
relation with superstring theory, it can be motivated from different sides.
First, it can be regarded as a toy model showing that certain nonlinear
field theories can produce particle-like solutions that can realize the
limiting curvature hypothesis in cosmological theories \cite{Soleng}.
Second, the behavior of logarithmic electrodynamics and BI electrodynamics
Lagrangians are the same for large values of nonlinear parameter $\beta $.
Third, from gauge/gravity duality point of view, the ratio of holographic
viscosity to entropy density is not affected by nonlinear terms rise from a
nonlinear electrodynamics in contrast with gravitational corrections \cite%
{vis}. Fourth, the values of important parameters of a holographic
superconductor system such as critical temperature and order parameter are
firmly sensitive to choice of electrodynamics \cite{ord,ord2}. Other studies
on the nonlinear electrodynamics have been carried out in \cite%
{Frad,Cal,Hendi,HendiSlowly,SheyHaj,SheyKaz,Shey1,DHSR}.

The above pointed out motivations are convincingly enough to satisfy one to
seek for the effects of logarithmic electrodynamics on the solutions. Till
now, exact rotating solutions of logarithmic electrodynamics in the context
of dilaton gravity has not been constructed. In this paper, we would like to
construct the rotating dilaton black branes in the presence of logarithmic
nonlinear electrodynamics and investigate their thermodynamics as well as
their thermal stability.

The outline of this paper is as follows. In the next section, we present the
basic field equations. In section \ref{brane}, we construct the rotating
dilaton black branes with a complete set of rotation parameters in all
higher dimensions and investigate their properties. In section \ref{Therm},
we study thermodynamics of the spacetime, by calculating the conserved and
thermodynamic quantities. In section \ref{stab}, we perform a stability
analysis and show that the dilaton creates an unstable phase for the
solutions. The last section is devoted to conclusions and discussions.

\section{Basic Field Equations}

\label{Field} We consider an $n$-dimensional action in which gravity is
coupled to a dilaton and a nonlinear electrodynamic field
\begin{eqnarray}
I &=&-\frac{1}{16\pi }\int_{\mathcal{M}}d^{n}x\sqrt{-g}\left( \mathcal{\ R}%
\text{ }-\frac{4}{n-2}(\nabla \Phi )^{2}-V(\Phi )+L(F,\Phi )\right)  \notag
\\
&&-\frac{1}{8\pi }\int_{\partial \mathcal{M}}d^{n-1}x\sqrt{-h}\Theta (h),
\label{Act}
\end{eqnarray}%
where the Lagrangian of the logarithmic nonlinear electrodynamics coupled to
the dilaton field (LNd) is chosen in the following form
\begin{equation}
L(F,\Phi )=-8\beta ^{2}e^{4\alpha \Phi /(n-2)}\ln \Bigg(1+\frac{e^{-8\alpha
\Phi /(n-2))}F^{2}}{8\beta ^{2}}\Bigg).  \label{LLND}
\end{equation}%
In action (\ref{Act}), $\mathcal{R}$ is the Ricci scalar curvature, $\Phi $
is the dilaton field, and $V(\Phi )$ is a potential for $\Phi $. The dilaton
parameter $\alpha $ determines the strength of coupling of the scalar and
LNd fields, $F^{2}=F^{\mu \nu }F_{\mu \nu }$, where $F_{\mu \nu }=\partial
_{\mu }A_{\nu }-\partial _{\nu }A_{\mu }$ is the electromagnetic tensor
field,$\ A_{\mu }$ is the vector potential, and $\beta $ is the nonlinear
parameter with dimension of mass. The last term in Eq. (\ref{Act}) is the
Gibbons-Hawking boundary term which is chosen such that the variational
principle is well-defined. The manifold $\mathcal{M}$ has metric $g_{\mu \nu
}$ and covariant derivative $\nabla _{\mu }$. $\Theta $ is the trace of the
extrinsic curvature $\Theta ^{ab}$ of any boundary(ies) $\partial \mathcal{M}
$ of the manifold $\mathcal{M}$, with induced metric(s) $h_{ab}$. In this
paper, we consider the action (\ref{Act}) with a Liouville type potential,
\begin{equation}
V(\Phi )=2\Lambda e^{4\alpha \Phi /(n-2)},  \label{v1}
\end{equation}%
where $\Lambda $ is a constant which may be referred to as the cosmological
constant, since in the absence of the dilaton field ($\Phi =0$) the action (%
\ref{Act}) reduces to the action of Einstein gravity in the presence of
nonlinear electrodynamics with cosmological constant. For later convenience,
we redefine it as $\Lambda =-(n-1)(n-2)/2l^{2}$, where $l$ is a constant
with dimension of length. The series expansion of (\ref{LLND}) for large $%
\beta $, leads to
\begin{equation*}
L_{\mathrm{LNd}}(F,\Phi )=-e^{-4\alpha \Phi /(n-2)}F^{2}+\frac{e^{-12\alpha
\Phi /(n-2)}F^{4}}{16\beta ^{2}}-\frac{e^{-20\alpha \Phi /(n-2)}F^{6}}{%
192\beta ^{4}}+\mathcal{O}\Bigg (\frac{1}{\beta ^{6}}\Bigg).
\end{equation*}%
For latter convenience we rewrite
\begin{equation*}
L_{\mathrm{LNd}}(F,\Phi )=-8\beta ^{2}e^{4\alpha \Phi /(n-2)}\mathcal{L}(Y),
\end{equation*}%
where we have defined
\begin{equation*}
\mathcal{L}(Y)=\ln (1+Y)\mathrm{,}
\end{equation*}%
\begin{equation*}
Y=\frac{e^{-8\alpha \Phi /(n-2)}F^{2}}{8\beta ^{2}}.
\end{equation*}%
By varying the action (\ref{Act}) with respect to the gravitational field $%
g_{\mu \nu }$, the dilaton field $\Phi $ and the gauge field $A_{\mu }$. We
find
\begin{eqnarray}
\mathcal{R}_{\mu \nu } &=&\frac{4}{n-2}\left( \partial _{\mu }\Phi \partial
_{\nu }\Phi +\frac{1}{4}g_{\mu \nu }V(\Phi )\right) +2e^{-4\alpha \Phi
/(n-2)}\partial _{Y}{\mathcal{L}}(Y)F_{\mu \eta }F_{\nu }^{\text{ }\eta }
\notag  \label{FE1} \\
&&-\frac{8\beta ^{2}}{n-2}e^{4\alpha \Phi /(n-2)}\left[ 2Y\partial _{Y}{%
\mathcal{L}}(Y)-{\mathcal{L}}(Y)\right] g_{\mu \nu },
\end{eqnarray}%
\begin{equation}
\nabla ^{2}\Phi =\frac{n-2}{8}\frac{\partial V}{\partial \Phi }-4\alpha
\beta ^{2}e^{4\alpha \Phi /(n-2)}\left[ 2{\ Y}\partial _{Y}{\mathcal{L}}(Y)-%
\mathcal{L}(Y)\right] ,  \label{FE2}
\end{equation}%
\begin{equation}
\nabla _{\mu }\left( e^{-4\alpha \Phi /(n-2)}\partial _{Y}{\mathcal{L}}%
(Y)F^{\mu \nu }\right) =0.  \label{FE3}
\end{equation}%
In the limiting case where $\beta \rightarrow \infty $, we have $\mathcal{L}%
(Y)=Y$. In this case the system of field equations (\ref{FE1})-(\ref{FE3})
restore the well-known equations of EMd gravity \cite%
{CHM,Shey3,Cai,yaz2,Clem}, as expected.

\subsection{FINITE ACTION IN CANONICAL AND GRAND-CANONICAL ENSEMBLES \label%
{Finite}}

In general, the total action $I$ given in Eq. (\ref{Act}) is divergent when
evaluated on a solution. One way of dealing with the divergences of the
action is adding some counterterms to the action (\ref{Act}). The
counterterms should contain a part which removes the divergence of the
gravity part of the action and a part for dealing with the divergence of the
matter action. Since the horizon of our solution is flat, the counterterm
which removes the divergence of the gravity part should be proportional to $%
\sqrt{-h}$. The counterterm for the matter part of the action in the
presence of the dilaton is given by
\begin{equation}
I_{\mathrm{ct}}=-\frac{1}{8\pi }\int_{\partial \mathcal{M}}\ d^{n-1}x \sqrt{%
-h}\left( \frac{n-2}{l_{ \mathrm{eff}}}\right)+I_{\mathrm{deriv}} ,
\end{equation}
where $l_{\mathrm{eff}}$ is given by (\ref{leff}) and $I_{\mathrm{deriv}}$
is a collection of terms involving derivatives of the boundary fields that
could involve the curvature tensor constructed from the boundary metric.
Since in our case the boundary is flat so $I_{\mathrm{deriv}}$ is zero on
the boundary. The variation of the total action $\left( I_{\mathrm{tot}}\
=I+I_{\mathrm{ct}}\right) $ about the solutions of the equations of motion
is
\begin{equation}
\delta I_{\mathrm{tot}}=\int d^{n-1}x S_{ab }\delta h^{ab }-\frac{1}{16\pi }%
\int d^{n-1}x\sqrt{-h} e^{-4\alpha \Phi/(n-2)} \partial_{Y}{\mathcal{L}}(Y)
n^{a }F_{ab }\delta A^{b},  \label{Ivar}
\end{equation}%
where
\begin{equation}
S_{ab }=\frac{\sqrt{-h}}{16\pi }\left\{\Theta _{ab}-\Theta h_{ab}+\frac{n-2}{%
l_{ \mathrm{eff}}}h_{ab} \right\}.
\end{equation}%
Equation (\ref{Ivar}) shows that the variation of the total action with
respect to $A^{a }$ will only give the equation of motion of the nonlinear
massless field $A^{a }$ provided the variation is at fixed nonlinear
massless gauge potential on the boundary. Thus, the total action, $I_{%
\mathrm{tot}}\ =I+I_{\mathrm{ct}}$, given in Eq. (\ref{Ivar}) is appropriate
for the grand-canonical ensemble, where $\delta A^{a }=0$ on the boundary.
But in the canonical ensemble, where the electric charge $\left[-e^{-4\alpha
\Phi/(n-2)} \partial_{Y}{\mathcal{L}}(Y) n^{a }F_{ab }\right]$ is fixed on
the boundary, the appropriate action is
\begin{equation}
I_{\mathrm{tot}}=I+I_{\mathrm{ct}}+\frac{1}{16\pi }\int_{\partial \mathcal{M}%
`}d^{n-1}x\ \sqrt[\ ]{-h}e^{-4\alpha \Phi/(n-2)} \partial_{Y}{\mathcal{L}}%
(Y) n^{a }F_{ab }\delta A^{b} .  \label{I3term}
\end{equation}
The last term in Eq. (\ref{I3term}) is the generalization of the boundary
term introduced by Hawking for linear electromagnetic field \cite{S. W.
Hawking grand} and the results of \cite{DSV,mhd2} for the nonlinear Lifshitz
black holes to the exponential nonlinear gauge field coupled to the dilaton
field. Thus, both in canonical and grand-canonical ensemble, the variation
of total action about the solutions of the field equations is
\begin{equation}
\delta I_{\mathrm{tot}}=\int d^{n-1}x S_{ab }\delta h^{ab }.
\end{equation}%
That is, the nonlinear gauge field is absent in the variation of the total
action both in canonical and grand-canonical ensembles.

In order to obtain the conserved charges of the spacetime, we use the
counterterm method \cite{BY,Mal} inspired by (A)dS/CFT correspondence. For
asymptotically AdS solutions this method works very well \cite{Mal}.
However, in our paper we have the scalar dilaton field with a Liouville
potential. It was argued that the presence of Liouville-type dilaton
potential, which is regarded as the generalization of the cosmological
constant, changes the asymptotic behavior of the solutions to be neither
asymptotically flat nor (A)dS. It has been shown that no dilaton dS or AdS
black hole solution exists with the presence of only one Liouville-type
dilaton potential \cite{MW}. But, as in the case of asymptotically AdS
spacetimes, according to the domain-wall/QFT (quantum field theory)
correspondence \cite{Sken}, there may be a suitable counterterm for the
stress-energy tensor which removes the divergences. In this paper, we deal
with the spacetimes with zero curvature boundary [$R_{abcd}(h)=0$], and
therefore the counterterm for the stress-energy tensor should be
proportional to $h^{ab}$. We find the finite stress-energy tensor in $n$%
-dimensional Einstein-dilaton gravity with Liouville-type in the form \cite%
{DHSR}
\begin{equation}
T^{ab}=\frac{1}{8\pi }\left[ \Theta ^{ab}-\Theta h^{ab}+\frac{n-2}{l_{%
\mathrm{eff}}}h^{ab}\right] ,  \label{Stres}
\end{equation}%
where $l_{\mathrm{eff}}$ is given by
\begin{equation}
l_{\mathrm{eff}}^{2}=\frac{(n-2)(\alpha ^{2}-n+1)}{2\Lambda }e^{-4\alpha
\Phi /(n-2)}.  \label{leff}
\end{equation}

In the particular case $\alpha =0$, the effective $l_{\mathrm{eff}}^{2}$ of
Eq. (\ref{leff}) reduces to $l^{2}=-(n-1)(n-2)/2\Lambda $ of the AdS
spacetimes. The first two terms in Eq. (\ref{Stres}) is the variation of the
action (\ref{Act}) with respect to $h_{ab}$, and the last term is the
counterterm which removes the divergences. One may note that the counterterm
has the same form as in the case of asymptotically AdS solutions with zero
curvature boundary, where $l$ is replaced by $l_{\mathrm{eff}}$. If we
choose the Killing vector field $\mathcal{\xi }$ on spacelike surface $%
\mathcal{B}$ in $\partial \mathcal{M}$ with metric $\sigma _{ij}$, then the
quasilocal conserved quantities may be obtained from the following relation
\cite{DHSR}
\begin{equation}
Q(\mathcal{\xi )}=\int_{\mathcal{B}}d^{n-2}x\sqrt{\sigma }T_{ab}n^{a}%
\mathcal{\ \xi }^{b},  \label{charge}
\end{equation}%
where $\sigma $ is the determinant of the boundary metric $\sigma _{ij}$ and
$n^{a}$ is the unit normal vector on the boundary $\mathcal{B}$. In our
case, the boundary $\mathcal{B}$ has two Killing vector fields timelike ($%
\partial /\partial t$) and rotational ($\partial /\partial \varphi $). The
corresponding conserved charges are the quasilocal mass and angular momentum
may be obtained as

\begin{eqnarray}
M &=&\int_{\mathcal{B}}d^{n-2}x \sqrt{\sigma }T_{ab}n^{a}\xi ^{b},
\label{Mastot} \\
J &=&\int_{\mathcal{B}}d^{n-2}x \sqrt{\sigma }T_{ab}n^{a}\varsigma ^{b}.
\label{Angtot}
\end{eqnarray}


\section{Rotating dilaton black branes in higher dimensions}

\label{brane} In this section, we would like to construct the rotating black
brane solutions of the field equations (\ref{FE1})-(\ref{FE3}) with $k$
rotation parameters. The number of independent rotation parameters for an $n$%
-dimensional localized object is equal to the number of Casimir operators,
which is $[(n-1)/2]\equiv k$, where $[x]$ is the integer part of $x$ \cite%
{SDRP}. The metric of $n$-dimensional rotating solution with cylindrical or
toroidal horizons and $k$ rotation parameters can be written as \cite%
{Lemos,awad}
\begin{eqnarray}
ds^{2} &=&-f(r)\left( \Xi dt-{{\sum_{i=1}^{k}}}a_{i}d\phi _{i}\right) ^{2}+%
\frac{r^{2}}{l^{4}}R^{2}(r){{\sum_{i=1}^{k}}}\left( a_{i}dt-\Xi l^{2}d\phi
_{i}\right) ^{2}  \notag \\
&&-\frac{r^{2}}{l^{2}}R^{2}(r){\sum_{i<j}^{k}}(a_{i}d\phi _{j}-a_{j}d\phi
_{i})^{2}+\frac{dr^{2}}{f(r)}+\frac{r^{2}}{l^{2}}R^{2}(r)dX^{2},  \notag \\
\Xi ^{2} &=&1+\sum_{i=1}^{k}\frac{a_{i}^{2}}{l^{2}},  \label{Met}
\end{eqnarray}
where $a_{i}$'s are $k$ rotation parameters. There are two unknown functions
$f(r)$ and $R(r)$ in the above metric which should be determined by solving
the field equations. The range of the angular coordinates are $0\leq \phi
_{i}\leq 2\pi $ and $dX^{2}$ is the Euclidean metric on the $(n-k-2)$%
-dimensional submanifold with volume $\Sigma _{n-k-2}$.

First of all, we integrate the electromagnetic field equation (\ref{FE3}).
The result is
\begin{eqnarray}
F_{{tr}} &=&\frac{2q\Xi e^{4\alpha \Phi /(n-2)}}{(rR(r))^{n-2}}\Bigg(1+\sqrt{%
1+\frac{q^{2}}{\beta ^{2}(rR(r))^{2n-4}}}\Bigg)^{-1},  \label{FtrE} \\
F_{\phi _{i}r} &=&-\frac{a_{i}}{\Xi }F_{tr},  \label{FprE}
\end{eqnarray}%
where $q$, is an integration constant related to the electric charge of the
brane. When $\beta \rightarrow \infty $, $F_{tr}$ reduces to the electric
field of $n$-dimensional black brane of Einstein-Maxwell-dilaton gravity
\cite{SDRP}
\begin{equation}
F_{tr}=\frac{q\Xi e^{4\alpha \Phi /(n-2)}}{(rR(r))^{n-2}}+O\left( \frac{1}{%
\beta ^{2}}\right) .  \label{FtrM}
\end{equation}%
In order to solve the system of equations (\ref{FE1}) and
(\ref{FE2}) for three unknown functions $f(r)$, $R(r)$ and $\Phi
(r)$, we make the ansatz \cite{SDRP}
\begin{equation}
R(r)=e^{2\alpha \Phi /(n-2)}.  \label{Rphi}
\end{equation}%
{In order to justify this choice for the metric function $R(r)$,
let us note that $R(r)$ is indeed added to the metric (\ref{Met})
in order to increase the degrees of freedom for obtaining
solutions in the presence of the dilaton field. Choosing $R(r)$ in
the form of Eq. (\ref{Rphi}), is an ansatz. However, it is chosen
such that in the absence of the dilaton field $\Phi=0$, we have
$R(r)=1$, as expected. With this ansatz, we are able to solve the
field equation, analytically.}

\begin{figure}[t]
\epsfxsize=7cm \centerline{\epsffile{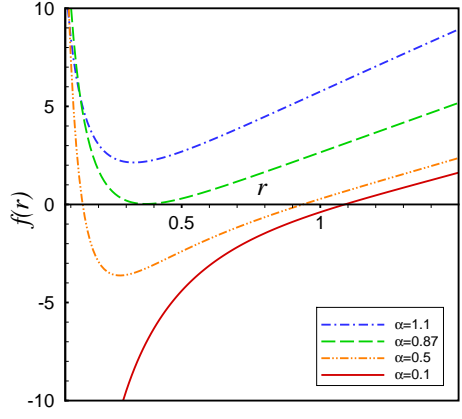}}
\caption{The behavior of $f(r)$ versus $r$ with $l=b=1$, $q=0.5$, $\Xi =1.25$%
, $n=5$, $\protect\beta =2$ and $m=1.5$.}
\label{fig1}
\end{figure}

Substituting (\ref{Rphi}), the electromagnetic fields (\ref{FtrE})- (\ref%
{FprE}) and the metric (\ref{Met}) into the field equations (\ref{FE1}) and (%
\ref{FE2}), one can obtain the following solutions
\begin{eqnarray}
f(r) &=&\frac{2(\alpha ^{2}+1)^{2}(\Lambda -4\beta ^{2})b^{\gamma }}{%
(n-2)(\alpha ^{2}-n+1)}r^{2-\gamma }-\frac{m}{r^{n-3-(n-2)\gamma /2}}  \notag
\\
&&-\frac{8\beta ^{2}(\alpha ^{2}+1)b^{\gamma }}{(n-2)r^{n-3-(n-2)\gamma /2}}%
\int r^{n(1-\frac{\gamma }{2})-2}\Bigg\{\sqrt{1+\eta }-\ln \bigg(\frac{\eta
}{2}\bigg)+\ln \Big(-1+\sqrt{1+\eta }\,\Big)\Bigg\}dr,  \label{f1}
\end{eqnarray}%
\begin{equation}
\Phi (r)=\frac{(n-2)\alpha }{2(\alpha ^{2}+1)}\ln \left(
c+\frac{b}{r}\right) , \label{phi}
\end{equation}%
{where } $c$  and   $b${\ are constant of integration. We find
that these solutions will fully satisfy the system of equations
(\ref{FE1}) and (\ref{FE2}) provided we choose $c=0$. Note that
$b$ has the dimension of [Length] to make the argument of
logarithmic function dimensionless}. In the above solutions
$\gamma =2\alpha ^{2}/(1+\alpha ^{2})$, and
\begin{equation}
\eta =\frac{q^{2}b^{(2-n)\gamma }}{\beta ^{2}r^{(n-2)(2-\gamma )}}.
\label{eta}
\end{equation}%
In the above expression, $m$ appears as an integration constant and is
related to the mass of the black hole. The integration of Eq. (\ref{f1}) can
be performed using the MATHEMATICA software. The resulting solution can be
written%
\begin{eqnarray}
f(r) &=&\frac{2(\Lambda -4\beta ^{2})(\alpha ^{2}+1)^{2}b^{\gamma }}{%
(n-2)(\alpha ^{2}-n+1)}r^{2-\gamma }-\frac{m}{r^{n-3-(n-2)\gamma /2}}  \notag
\\
&&+\frac{8\beta ^{2}(\alpha ^{2}+1)^{2}}{(\alpha ^{2}-n+1)^{2}}b^{\gamma
}r^{2-\gamma }\Bigg\{1-{}_{2}F_{1}\Bigg(\Bigg[\frac{-1}{2}\,,\frac{\alpha
^{2}-n+1}{2n-4}\Bigg]\,,\Bigg[\frac{\alpha ^{2}+n-3}{2n-4}\Bigg]\,,-\eta %
\Bigg)\Bigg\}  \notag \\
&&+\frac{8\beta ^{2}(\alpha ^{2}+1)^{2}}{(n-2)(\alpha ^{2}-n+1)}\,b^{\gamma
}r^{2-\gamma }\Bigg\{\sqrt{1+\eta }-\ln \left( {\frac{\eta }{2}}\right) +\ln %
\Big(-1+\sqrt{1+\eta }\,\Big)\Bigg\},  \label{f2}
\end{eqnarray}%
where ${}_{2}F_{1}([a,b],[c],z)$ is the hypergeometric function \cite%
{Lambert}. It is worth mentioning that the solutions are ill-defined for $%
\alpha =\sqrt{n-1}$. We expect that for large $\beta $, the function $f(r)$
reduces to the $n$-dimensional charged rotating dilaton black brane
solutions given in Ref. \cite{SDRP}. Indeed, if we expand Eq. (\ref{f2}) for
large $\beta $, we arrive at%
\begin{eqnarray}
f(r) &=&\frac{2\Lambda (\alpha ^{2}+1)^{2}}{(n-2)(\alpha ^{2}-n+1)}b^{\gamma
}r^{2-\gamma }-\frac{m}{r^{n-3-(n-2)\gamma /2}}  \notag  \label{f3} \\
&&+\frac{2q^{2}(\alpha ^{2}+1)^{2}b^{-(n-3)\gamma }}{(n-2)(\alpha
^{2}+n-3)r^{(n-3)(2-\gamma )}}-\frac{q^{4}(\alpha
^{2}+1)^{2}b^{-(2n-5)\gamma }}{4\beta ^{2}(n-2)(\alpha
^{2}+3n-7)r^{(2n-5)(2-\gamma )}}+\mathcal{O}\Bigg(\frac{1}{\beta ^{4}}\Bigg).
\end{eqnarray}%
Setting $\alpha =\gamma =0$ in (\ref{f3}), we reach%
\begin{equation*}
f(r)=\frac{r^{2}}{l^{2}}-\frac{m}{r^{n-3}}+\frac{2q^{2}}{(n-2)(n-3)r^{2n-6}}-%
\frac{1}{4\beta ^{2}(n-2)(3n-7)}\frac{q^{4}}{r^{4n-10}}+\mathcal{O}\Bigg(%
\frac{1}{\beta ^{4}}\Bigg).
\end{equation*}%
The last term in the right hand side of the above expression is the leading
nonlinear correction to the AdS black brane with dilaton field. In the
absence of a nontrivial dilaton ($\alpha =\gamma =0$), the above solutions
reduce to the asymptotically AdS charged rotating black brane solutions of
Einstein gravity in the presence of EN electrodynamics \cite{Hendi}.
Finally, in the limit $\beta ^{2}\rightarrow \infty $ and $\alpha =0=\gamma $%
, the solution given by Eq. (\ref{f3}) has the form of the asymptotically
AdS black brane solutions \cite{awad,Deh4}. Figs. (\ref{fig1}) and (\ref%
{fig2}) depict the behavior of $f(r)$ given by Eq. (\ref{f2}) for different $%
\alpha $'s and $\beta $'s respectively.
\begin{figure}[t]
\epsfxsize=7cm \centerline{\epsffile{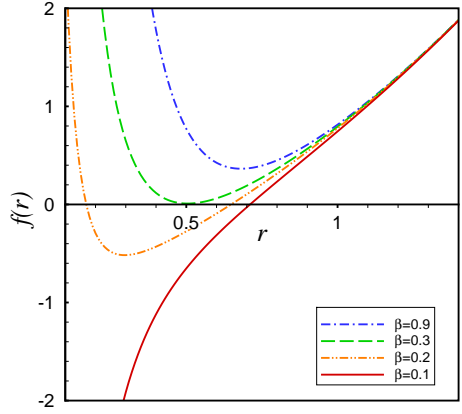}}
\caption{The behavior of $f(r)$ versus $r$ with $l=b=1$, $q=0.8$, $\Xi =1.25$%
, $n=5$, $\protect\alpha =0.2$ and $m=0.5$.}
\label{fig2}
\end{figure}

\subsection{Asymptotic behavior of the spacetime}

Next, we study the geometry of this spacetime. For this purpose, we first
seek for the curvature singularities in the presence of dilaton and
nonlinear electrodynamic fields. It is a matter of calculation to show that
the Ricci scalar and the Kretschmann invariant behave as
\begin{gather}
\lim_{r\longrightarrow 0^{+}}R=\infty ,  \label{Rorigin} \\
\lim_{r\longrightarrow 0^{+}}R_{\mu \nu \rho \sigma }R^{\mu \nu \rho \sigma
}=\infty ,  \label{RRorigin}
\end{gather}%
which indicate that there is an essential singularity at $r=0$. In order to
study the asymptotic behavior of the solutions, we expand the metric
function $f(r)$ for $r\rightarrow \infty $ limit. We find
\begin{equation}
\lim_{r\longrightarrow \infty }f(r)=\frac{2\Lambda (\alpha ^{2}+1)^{2}}{%
(n-2)(\alpha ^{2}-n+1)}b^{\gamma }r^{2-\gamma }.  \label{fasymp}
\end{equation}%
Let us note that in the absence of the dilaton field ($\alpha =0=\gamma $),
the metric function becomes
\begin{equation}
\lim_{r\longrightarrow \infty }f(r)=-\frac{2\Lambda r^{2}}{(n-1)(n-2)},
\end{equation}%
which describes an asymptotically AdS ($\Lambda <0$) or dS ($\Lambda >0$)
spacetimes. However, as one can see from Eq. (\ref{fasymp}), in the presence
of the dilaton field, the asymptotic behavior is neither flat nor (A)dS. For
example, taking $\alpha =\sqrt{2}$, $n=6$ and $b=1$, we have
\begin{equation}
\lim_{r\longrightarrow \infty }f(r)=-\frac{3\Lambda }{2}r^{2/3}.
\label{fasymp3}
\end{equation}%
Clearly, the metric function (\ref{fasymp3}) is neither flat nor (A)dS. This
is consistent with the argument given in \cite{MW}, which states that no
dilaton dS or AdS black hole solution exists with the presence of only one
or two Liouville-type dilaton potential. It is important to note that this
asymptotic behavior is not due to the nonlinear nature of the electrodynamic
field, since as $r\rightarrow \infty $ the effects of the nonlinearity
disappear. Besides, from the dilaton field (\ref{phi}) we see that as $%
r\rightarrow \infty $, the dilaton field does not vanishes, while in case of
asymptotic flat or (A)dS we expect to have $\lim_{r\longrightarrow \infty
}\Phi (r)=0.$ Indeed, by solving the field equation (\ref{FE2}) we find
\begin{equation}
\Phi (r)=\frac{(n-2)\alpha }{2(\alpha ^{2}+1)}\ln \left( c+\frac{b}{r}%
\right) ,
\end{equation}%
however, the system of equation (\ref{FE1})-(\ref{FE3}) will be fully
satisfied provided we choose $c=0$. From the above arguments we conclude
that the asymptotic behavior of the obtained solutions is neither flat nor
(A)dS.

\section{Thermodynamics of black branes}

\label{Therm} It is easy to show that the metric given by (\ref{Met}) and (%
\ref{f2}) has both Killing and event horizons \cite{SDRP}. The Killing
horizon is a null surface whose null generators are tangent to a Killing
field. It is easy to see that the Killing vector
\begin{equation}
\chi =\partial _{t}+{{{\sum_{i=1}^{k}}}}\Omega _{i}\partial _{\phi _{i}},
\label{Kil}
\end{equation}%
is the null generator of the event horizon, where $\Omega _{i}$ is the $i$th
component of angular velocity of the outer horizon which may be obtained by
analytic continuation of the metric. The Hawking temperature and the angular
velocities of the outer event horizon can be obtained as
\begin{eqnarray}
T_{+} &=&\frac{f^{\text{ }^{\prime }}(r_{+})}{4\pi \Xi }=-\frac{\alpha ^{2}+1%
}{4\pi \Xi }r_{+}^{1-\gamma }\Bigg\{\frac{2(\Lambda -4\beta ^{2})b^{\gamma }%
}{(n-2)}  \notag  \label{Tem2} \\
&&+\frac{8\beta ^{2}b^{\gamma }}{n-2}\Bigg[\sqrt{1+\eta _{+}}-\ln \bigg(%
\frac{\eta _{+}}{2}\bigg)+\ln \Big(-1+\sqrt{1+\eta _{+}}\Big)\Bigg]\Bigg\},
\label{Temp} \\
\Omega _{i} &=&\frac{a_{i}}{\Xi l^{2}},  \label{Om1}
\end{eqnarray}%
where $\eta _{+}=\eta (r=r_{+})$ and we have used $f(r_{+})=0$ for deleting $%
m$. For large $\beta $, we can expand $T_{+}$ and arrive at the temperature
of the higher dimensional black branes in EMd gravity \cite{SDRP}
\begin{equation}
T_{+}=-\frac{\Lambda (\alpha ^{2}+1)b^{\gamma }}{2\pi \Xi (n-2)}%
r_{+}^{1-\gamma }-\frac{q^{2}(\alpha ^{2}+1)b^{-\gamma (n-3)}}{2\pi \Xi (n-2)%
}r_{+}^{5-2n-3\gamma +n\gamma }+\mathcal{O}\Bigg(\frac{1}{\beta ^{2}}\Bigg).
\label{Texp}
\end{equation}

The mass and angular momentum of the black branes ($\alpha <\sqrt{n-1}$) can
be calculated through the use of Eqs. (\ref{Mastot}) and (\ref{Angtot}).
Denoting the volume of the hypersurface boundary at constant $t$ and $r$ by $%
V_{n-2}=(2\pi )^{k}\Sigma _{n-k-2}$, the mass and angular momentum per unit
volume $V_{n-2}$ of the black branes can be obtained as
\begin{equation}
{M}=\frac{b^{(n-2)\gamma /2}}{16\pi l^{n-3}}\left\{ \frac{(n-1-\alpha
^{2})\Xi ^{2}+\alpha ^{2}-1}{1+\alpha ^{2}}\right\} m,  \label{Mass}
\end{equation}%
\begin{equation}
J_{i}=\frac{b^{(n-2)\gamma /2}}{16\pi l^{n-3}}\left( \frac{n-1-\alpha ^{2}}{%
1+\alpha ^{2}}\right) \Xi ma_{i}.  \label{Angmom}
\end{equation}%
{Note that, in order to avoid repeating the factor }$V_{n-2}${%
, we calculate, in this paper, the mass }$M$ {and extensive quantities such as angular momentum }$%
J_{i}$, {entropy }$S$ and charge $Q${\ appearing in first law
of thermodynamics per unit volume. }For the static case where $%
a_{i}=0$ ($\Xi =1$), the angular momentum per unit volume vanishes, and
therefore $a_{i}$'s are the rotational parameters of the black branes.

Black hole entropy typically satisfies the so called area law of the entropy
\cite{Beck}. This near universal law applies to almost all kinds of black
holes and black branes in Einstein gravity \cite{hunt}. It is easy to show
that the entropy per unit volume $V_{n-2}$ of the black brane can be written
as
\begin{equation}
{S}=\frac{\Xi b^{(n-2)\gamma /2}r_{+}^{(n-2)(1-\gamma /2)}}{4l^{n-3}},
\label{Entropy}
\end{equation}%
The electric charge per unit volume $V_{n-1}$ can be found by calculating
the flux of the electric field at infinity, yielding
\begin{equation}
{Q}=-\frac{1}{4\pi V_{n-1}}\int_{\Sigma }\nabla _{\mu }\left( \partial _{Y}{%
\mathcal{L}}(Y)F^{\mu \nu }\right) dS_{\nu }=-\frac{1}{8\pi V_{n-1}}%
\oint_{\partial \Sigma }\partial _{Y}{\mathcal{L}}(Y)F^{\mu \nu }dS_{\mu \nu
}=\frac{\Xi q}{4\pi l^{n-3}},  \label{Charge}
\end{equation}%
where the volume is replaced by an arbitrary spacelike hypersurface $\Sigma $
(partial Cauchy surface) with boundary $\partial \Sigma $. In addition, the
volume element on $\Sigma $ is a non-spacelike covector ($1$-form) $dS_{\nu
} $ and $dS_{\mu \nu }$ is the area element of $\partial \Sigma $. We should
note that for linear Maxwell case ($\beta \longrightarrow \infty $), one
obtains $\partial _{Y}{\mathcal{L}}(Y)=-1$.

The electric potential $U$, measured at infinity with respect to the
horizon, is defined by
\begin{equation}
U=A_{\mu }\chi ^{\mu }\left\vert _{r\rightarrow \infty }-A_{\mu }\chi ^{\mu
}\right\vert _{r=r_{+}},  \label{Pot1}
\end{equation}%
where $\chi $ is the null generator of the horizon given by Eq. (\ref{Kil}).
One can easily show that the vector potential $A_{\mu }$ corresponding to
the electromagnetic tensor (\ref{FtrE}) and (\ref{FprE}) can be written as
\begin{eqnarray}
A_{\mu } &=&\left( \Xi \delta _{\mu }^{t}-a_{i}\delta _{\mu }^{i}\right)
\times \frac{q(\alpha ^{2}+1)b^{(4-n)\gamma /2}}{\alpha ^{2}+n-3}%
r^{3-n-(4-n)\gamma /2}{} \\
&&\times {}_{3}F_{2}\Bigg(\Bigg[\frac{1}{2}\,,1\,,\frac{3-n-\alpha ^{2}}{4-2n%
}\Bigg]\,,\Bigg[2\,,\frac{7-3n-\alpha ^{2}}{4-2n}\Bigg]\,,-\eta \Bigg),
\notag
\end{eqnarray}%
where ${}_{3}F_{2}$ is the hypergeometric function and we have set the
constant of integration equal to zero. Therefore, the electric potential may
be obtained as
\begin{eqnarray}
U &=&\frac{q(\alpha ^{2}+1)b^{(4-n)\gamma /2}}{\Xi (\alpha ^{2}+n-3)}%
r_{+}^{3-n-(4-n)\gamma /2}{} \\
&&\times {}_{3}F_{2}\Bigg(\Bigg[\frac{1}{2}\,,1\,,\frac{3-n-\alpha ^{2}}{4-2n%
}\Bigg]\,,\Bigg[2\,,\frac{7-3n-\alpha ^{2}}{4-2n}\Bigg]\,,-\eta _{+}\Bigg).
\notag  \label{Pot}
\end{eqnarray}%
Now, we are in a position to verify the first law of thermodynamics. In
order to do this, we obtain the mass $M$ as a function of extensive
quantities $S$, $\mathbf{J}$ and $Q$. Using the expression for the mass, the
angular momenta, the entropy, and the charge given in Eqs. (\ref{Mass}), (%
\ref{Angmom}), (\ref{Entropy}), (\ref{Charge}) and the fact that $f(r_{+})=0$%
, one can obtain a Smarr-type formula as
\begin{equation}
M(S,\mathbf{J},Q)=\frac{\left[ (n-1-\alpha ^{2})Z+\alpha ^{2}-1\right]
\mathbf{J}}{l(n-1-\alpha ^{2})\sqrt{Z(Z-1)}},  \label{Smar}
\end{equation}%
where $\mathbf{J}=\sqrt{\sum_{i}^{k}{J_{i}}^{2}}$, and $Z=\Xi ^{2}$ is the
positive real root of the following equation
\begin{eqnarray}
&&\mathbf{J}+\frac{\beta ^{2}l^{4-n}(\alpha ^{2}+1)}{2\pi (n-2)(n-1-\alpha
^{2})}b^{\alpha ^{2}}\sqrt{Z(Z-1)}\left( \frac{4Sl^{n-3}}{\sqrt{Z}}\right) ^{%
\frac{n-1-\alpha ^{2}}{n-2}}\bigg\{(n-1-\alpha ^{2})\bigg [\ln \left( -1+%
\sqrt{1+\zeta }\right) -\ln \left( \frac{\zeta }{2}\right) +\sqrt{1+\zeta }%
\bigg]  \notag \\
&&+(n-2)\ {}_{2}F_{1}\left( \left[ -\frac{1}{2}\,,\frac{\alpha ^{2}-n+1}{2n-4%
}\right] \,,\left[ \frac{\alpha ^{2}+n-3}{2n-4}\right] \,,-\zeta \right) +%
\frac{(n-1)(n-2)}{8l^{2}\beta ^{2}}(\alpha ^{2}-n+1)+\alpha ^{2}-2n+3\bigg\}%
=0.
\end{eqnarray}%
where $\zeta =\pi ^{2}Q^{2}/\left( S^{2}\beta ^{2}\right) $. We can regard
the parameters $S$, $\mathbf{J}$, and $Q$ as a complete set of extensive
parameters for the mass $M(S,\mathbf{J},Q)$ and define the intensive
parameters conjugate to $S$, $\mathbf{J}$ and $Q$. This parameters are,
respectively, the temperature, the angular velocities, and the electric
potential, which are defined as
\begin{equation}
T=\left( \frac{\partial M}{\partial S}\right) _{J,Q},\ \ \Omega _{i}=\left(
\frac{\partial M}{\partial J_{i}}\right) _{S,Q},\ \ U=\left( \frac{\partial M%
}{\partial Q}\right) _{S,\mathbf{J}}.  \label{Dsmar}
\end{equation}%
Numerical calculations show that the intensive quantities calculated by Eq. (%
\ref{Dsmar}) coincide with Eqs. (\ref{Temp}), (\ref{Om1}) and (\ref{Pot}).
Thus, these thermodynamics quantities satisfy the first law of
thermodynamics
\begin{equation}
dM=TdS+{{{\sum_{i=1}^{k}}}}\Omega _{i}d{J}_{i}+Ud{Q}.
\end{equation}


\section{Thermal Stability of the black branes in canonical and
grand-canonical ensembles}

\label{stab}
\begin{figure}[t]
\epsfxsize=7cm \centerline{\epsffile{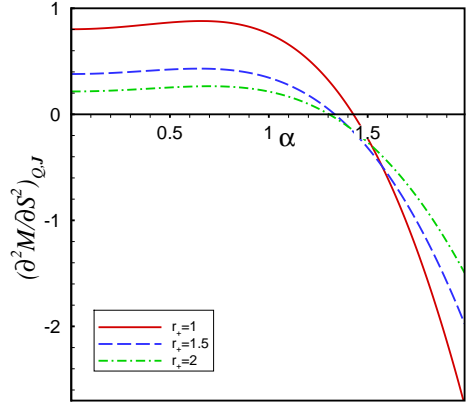}}
\caption{The behavior of $(\partial ^{2}M/\partial S^{2})_{Q,\mathbf{J}}$
versus $\protect\alpha $ with $l=b=1$, $q=0.8$, $\Xi =1.25$, $n=5$ and $%
\protect\beta =2$.}
\label{fig3}
\end{figure}
\begin{figure}[t]
\epsfxsize=7cm \centerline{\epsffile{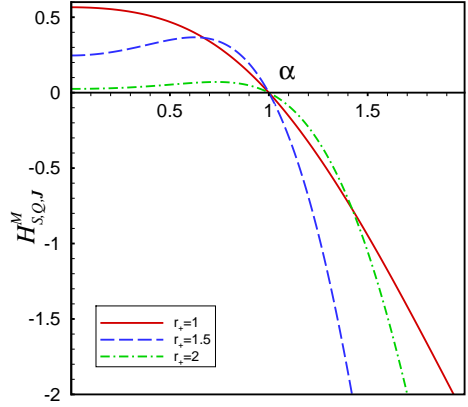}}
\caption{The behavior of $\mathbf{H}_{SQ\mathbf{J}}^{M}$ versus $\protect%
\alpha $ with $l=b=1$, $q=0.8$, $\Xi =1.25$, $n=5$ and $\protect\beta =2$.
Note that the curve corresponding to $r_{+}=1$ rescaled by a factor $10^{-1}$%
.}
\label{fig4}
\end{figure}
\begin{figure}[t]
\epsfxsize=7cm \centerline{\epsffile{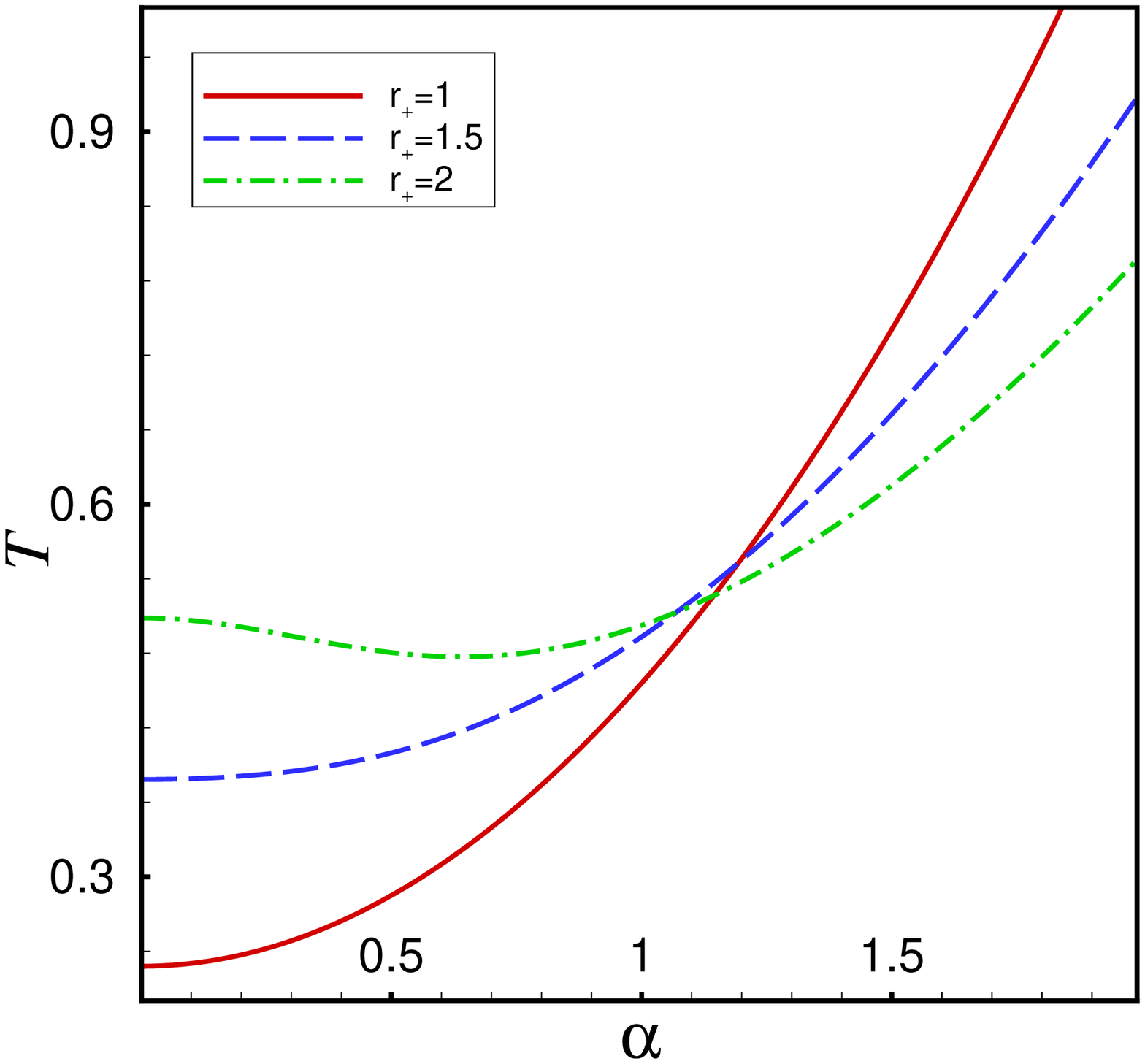}}
\caption{The behavior of $T$ versus $\protect\alpha $ with $l=b=1$, $q=0.8$,
$\Xi =1.25$, $n=5$ and $\protect\beta =2$.}
\label{fig5}
\end{figure}
\begin{figure}[t]
\epsfxsize=7cm \centerline{\epsffile{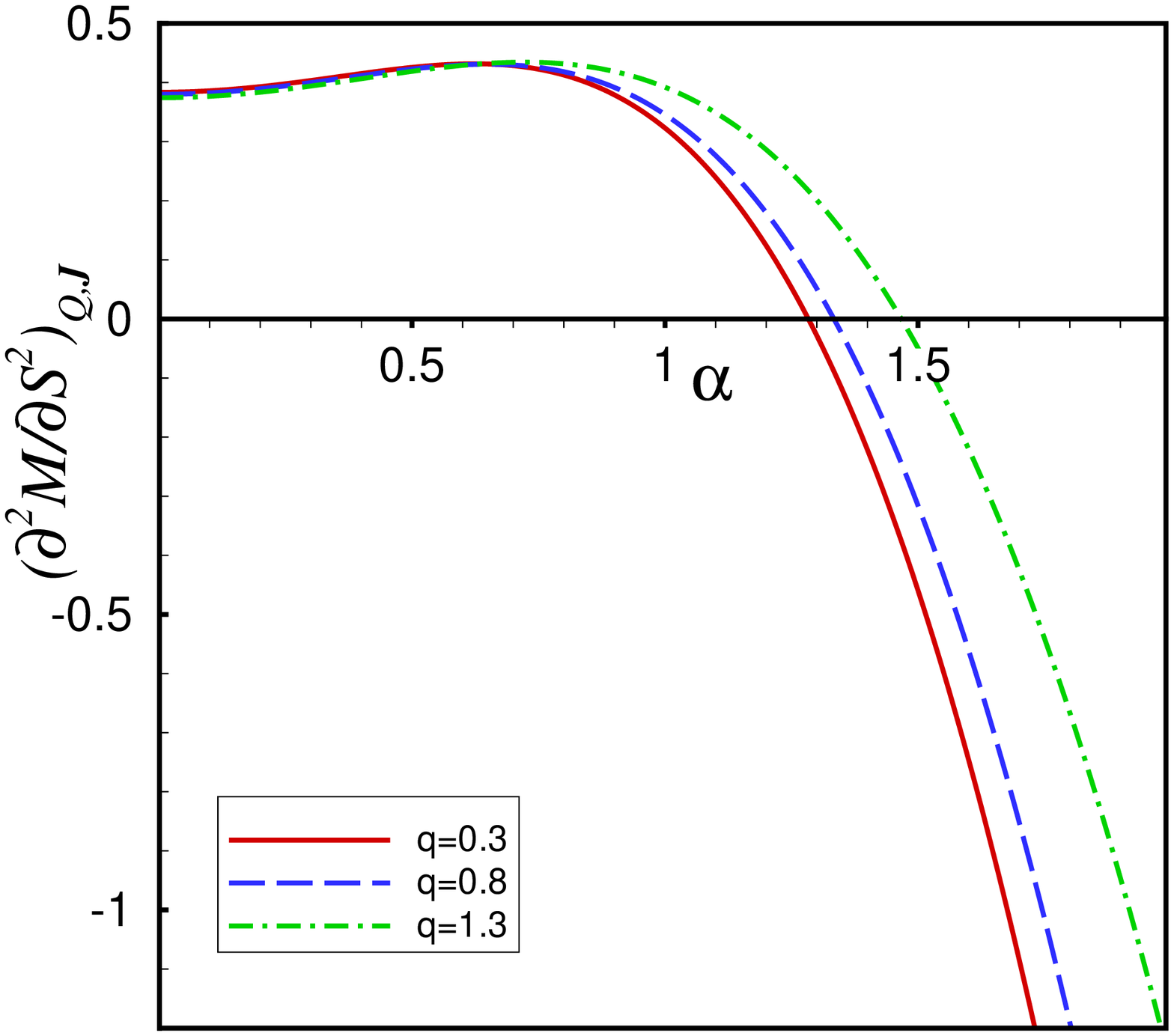}}
\caption{The behavior of $(\partial ^{2}M/\partial S^{2})_{Q,\mathbf{J}}$
versus $\protect\alpha $ with $l=b=1$, $r_{+}=1.5$, $\Xi =1.25$, $n=5$ and $%
\protect\beta =2$.}
\label{fig6}
\end{figure}
\begin{figure}[t]
\epsfxsize=7cm \centerline{\epsffile{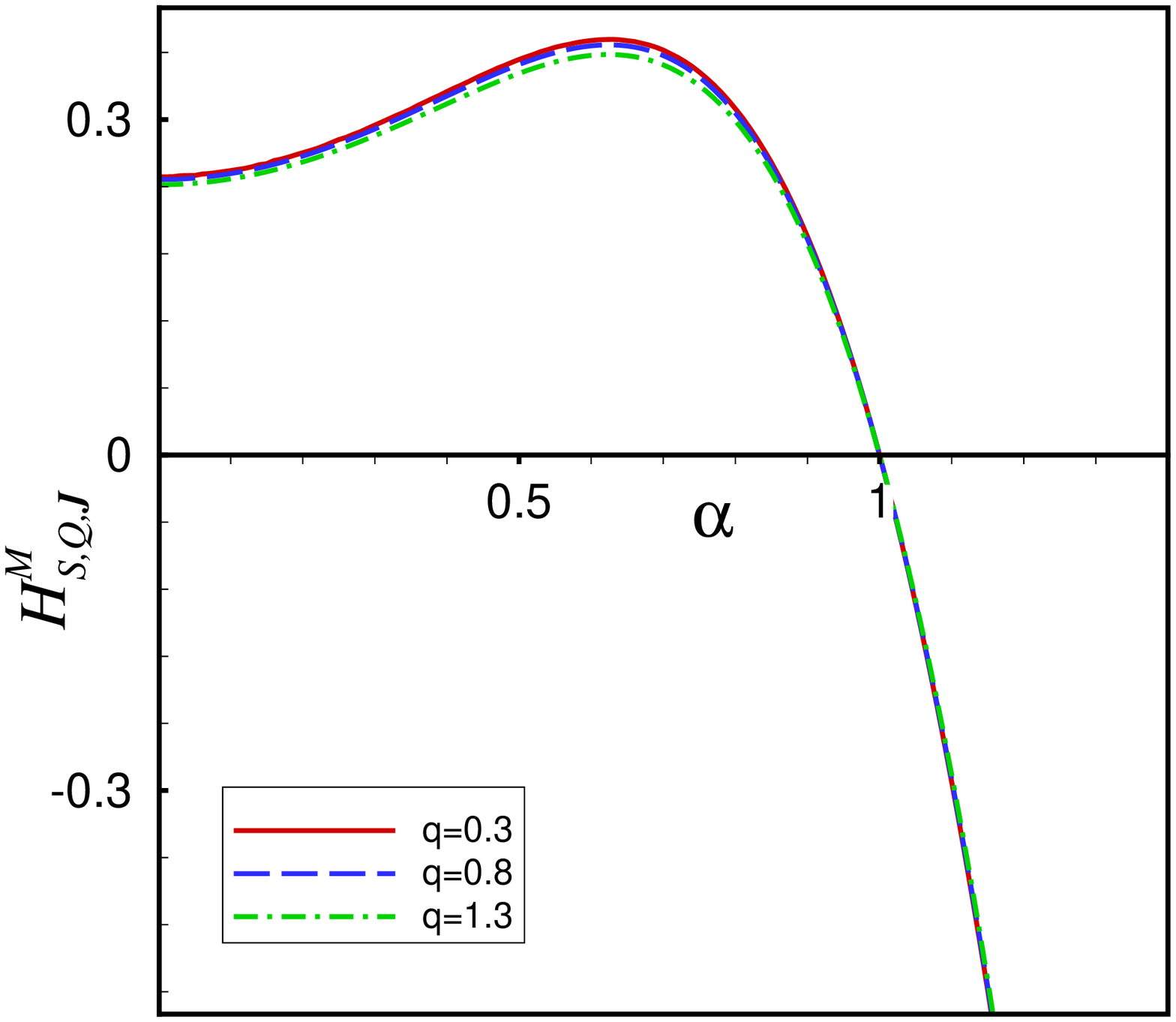}}
\caption{The behavior of $\mathbf{H}_{SQ\mathbf{J}}^{M}$ versus $\protect%
\alpha $ with $l=b=1$, $r_{+}=1.5$, $\Xi =1.25$, $n=5$ and $\protect\beta =2$%
.}
\label{fig7}
\end{figure}
\begin{figure}[t]
\epsfxsize=7cm \centerline{\epsffile{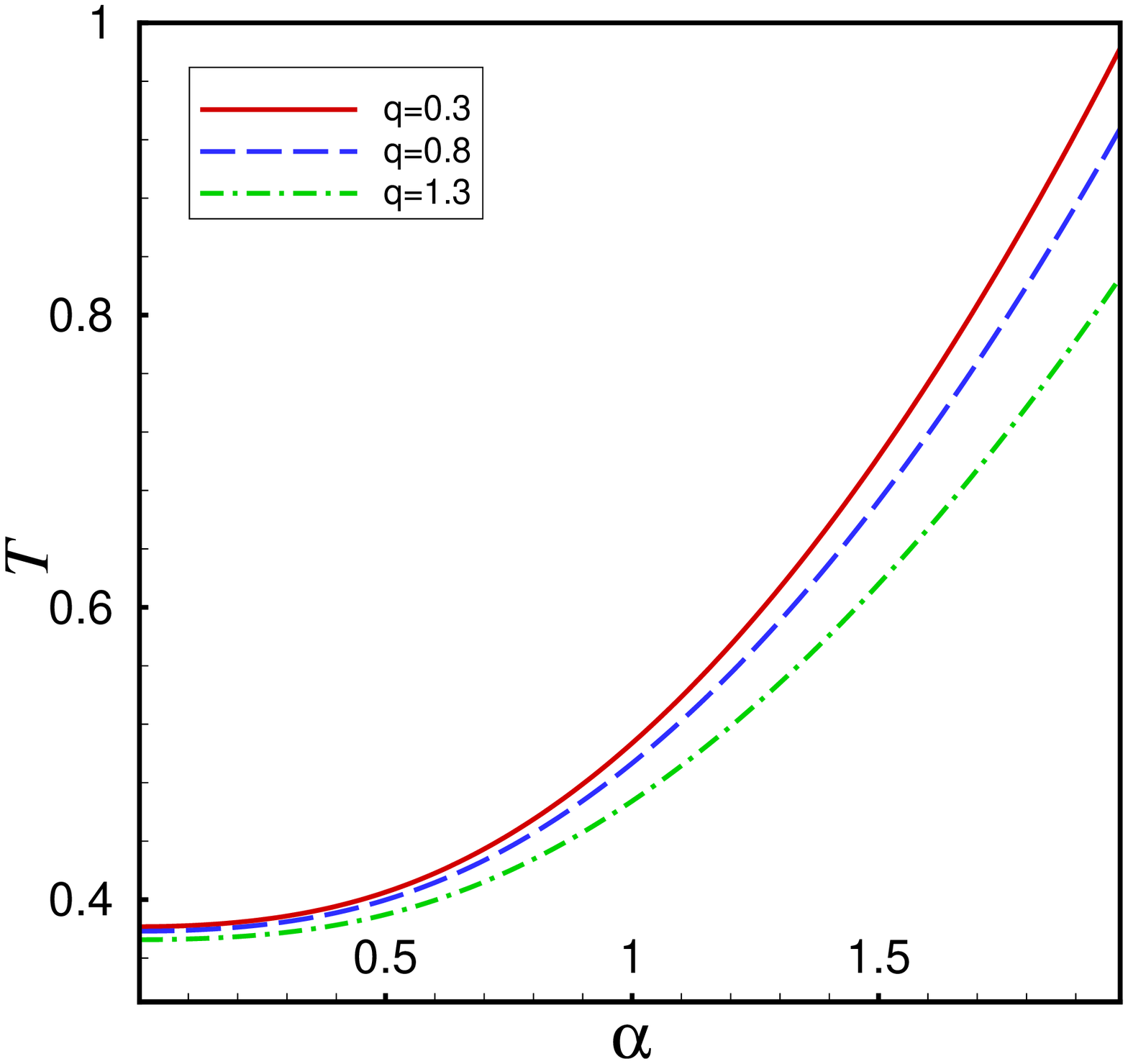}}
\caption{The behavior of $T$ versus $\protect\alpha $ with $l=b=1$, $%
r_{+}=1.5$, $\Xi =1.25$, $n=5$ and $\protect\beta =2$.}
\label{fig8}
\end{figure}
\begin{figure}[t]
\epsfxsize=7cm \centerline{\epsffile{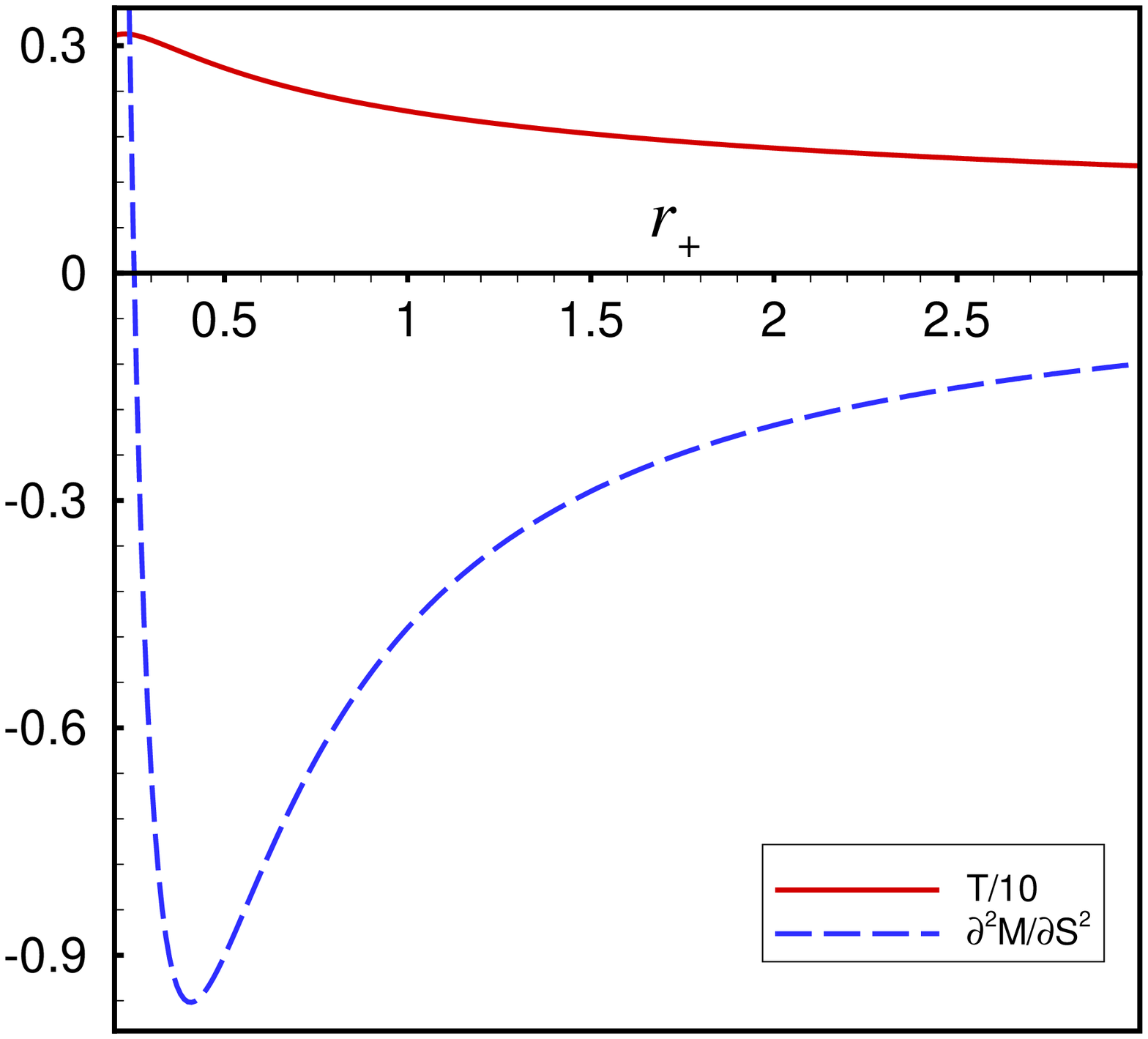}}
\caption{The behaviors of $10^{-1}T$ and $(\partial ^{2}M/\partial S^{2})_{Q,%
\mathbf{J}}$ versus $r_{+}$ with $l=1$, $b=2$, $\protect\alpha =1.5$, $\Xi
=1.25$, $n=5$, $\protect\beta =5$ and $q=1.1$.}
\label{fig9}
\end{figure}
\begin{figure}[t]
\epsfxsize=7cm \centerline{\epsffile{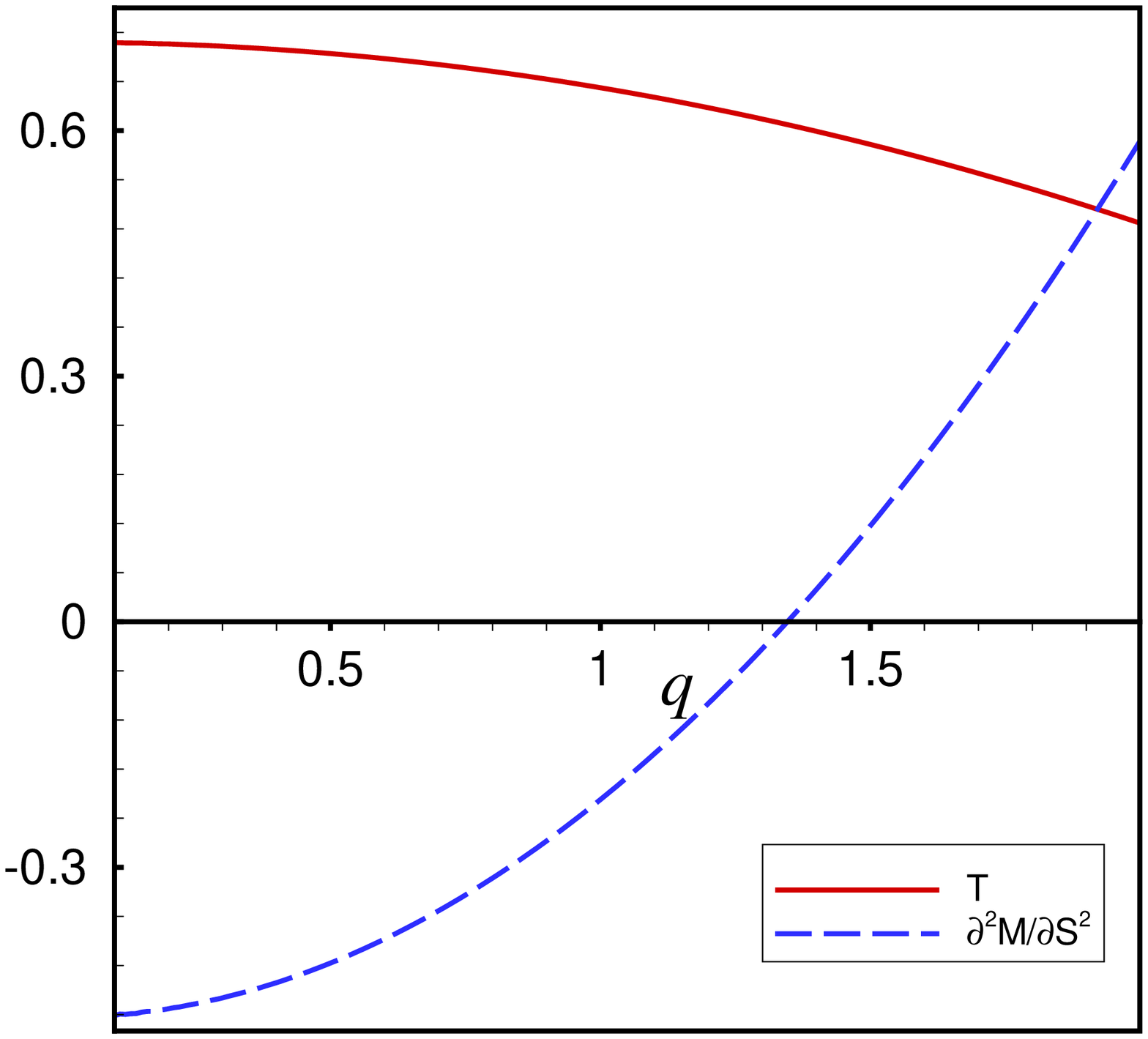}}
\caption{The behaviors of $T$ and $(\partial ^{2}M/\partial S^{2})_{Q,%
\mathbf{J}}$ versus $q$ with $l=b=1$, $\protect\alpha =1.5$, $\Xi =1.25$, $%
n=5$, $\protect\beta =5$ and $r_{+}=1.5$.}
\label{fig10}
\end{figure}
\begin{figure}[t]
\epsfxsize=7cm \centerline{\epsffile{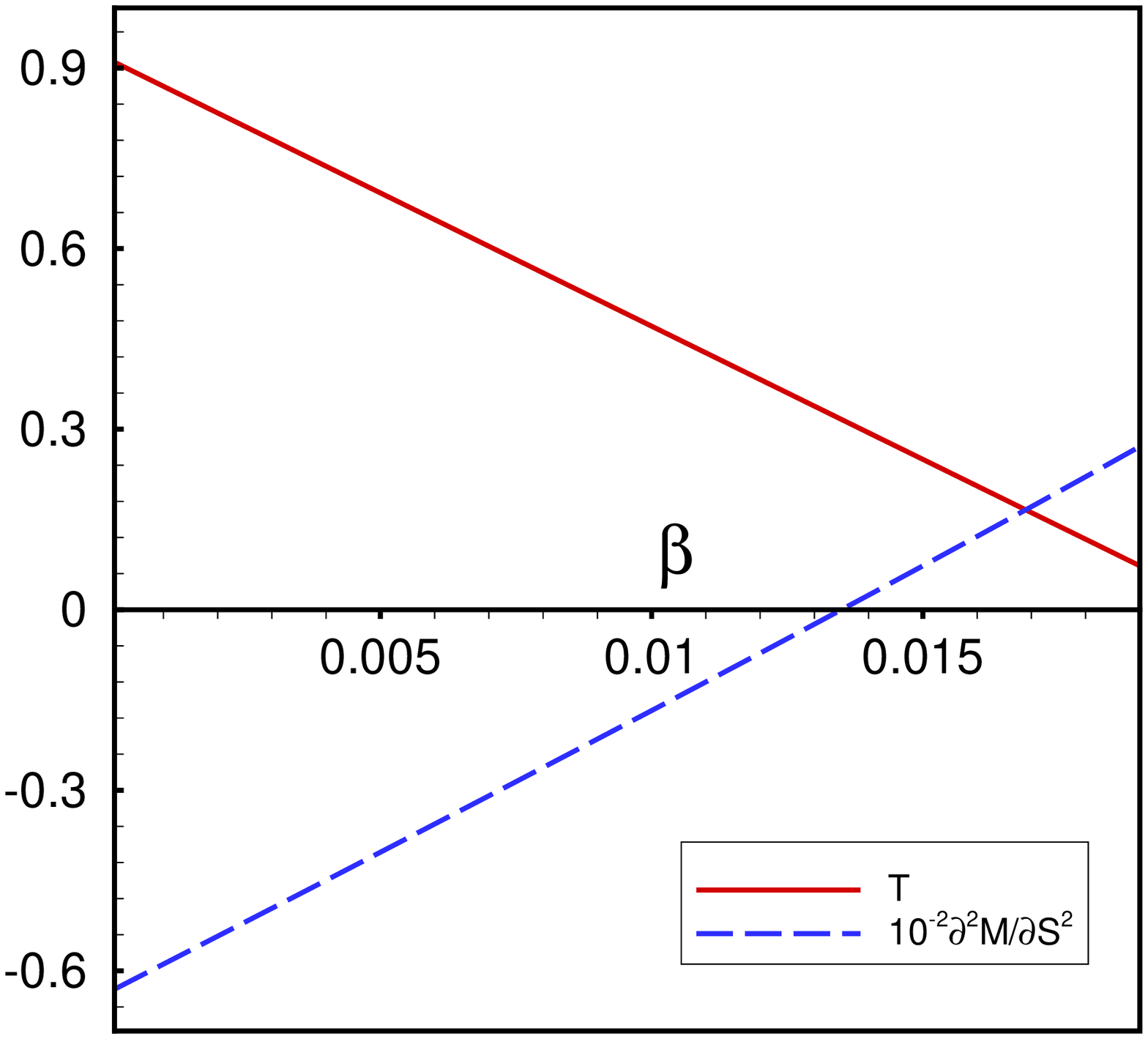}}
\caption{The behaviors of $T$ and $10^{-2}(\partial ^{2}M/\partial S^{2})_{Q,%
\mathbf{J}}$ versus $\protect\beta $ with $l=1$, $b=0.5$, $\protect\alpha =3$%
, $\Xi =1.25$, $n=6$, $q=10$ and $r_{+}=1$.}
\label{fig11}
\end{figure}

In this section, we intend to investigate thermal stability of our
nonlinearly charged rotating black brane solutions in both canonical and
grand-canonical ensembles. We know that the entropy of a thermally stable
system is at local maximum. The aim of thermal stability analysis is to find
the situations under which the system is stable thermally i.e. its entropy
is a local maximum. Therefore, the stability of charged rotating black brane
is studied in terms of entropy $S(M,Q,\mathbf{J})$. However, the thermal
stability can also be discussed in terms of internal energy. When the
entropy is at local maximum, the internal energy is at local minimum. Hence,
we can equivalently analyse thermal stability in terms of Legendre
transformation of entropy namely internal energy $M(S,Q,\mathbf{J})$. This
analysis is commonly done by studying the determinant of the Hessian matrix
of $M(S,Q,\mathbf{J})$\ with respect to its extensive variables $X_{i}$, $%
\mathbf{H}_{X_{i}X_{j}}^{M}=\left[ \partial ^{2}M/\partial X_{i}\partial
X_{j}\right] $\ \cite{Cal1,Gub}. The positivity of $\mathbf{H}%
_{X_{i}X_{j}}^{M}$ shows that the system is thermally stable. The number of
thermodynamic variables depends on the ensemble in which the system is
studied. For instance, in canonical ensemble where the charge and angular
momenta are fixed, the entropy is the only variable and consequently $%
\mathbf{H}_{X_{i}X_{j}}^{M}$ reduces to $(\partial ^{2}M/\partial S^{2})_{Q,%
\mathbf{J}}$. Thus, in this ensemble, the positivity of $(\partial
^{2}M/\partial S^{2})_{Q,\mathbf{J}}$ is sufficient to ensure the thermal
stability of course in the ranges the temperature $T$ is positive as well.
In grand-canonical ensemble $Q$\ and $\mathbf{J}$\ are no longer fixed.

Since the presence of charge does not change stable solutions to unstable
ones \cite{Deh4}, we first study thermal stability for uncharged case i.e. $%
q\rightarrow 0$. In this case
\begin{equation}
\left( \frac{\partial ^{2}M}{\partial S^{2}}\right) _{\mathbf{J}}=\frac{%
(n-1)\left( {\alpha }^{2}+1\right) \left[ ({\Xi }^{2}-1)(n-2{\alpha }^{2})+{%
\Xi }^{2}\left( 1-{\alpha }^{2}\right) \right] }{{\pi \Xi }^{2}{l}^{5-n}{b}^{%
{\left( n-4\right) \gamma /2}}\left[ ({\alpha }^{2}+n-3){\Xi }^{2}+1-{\alpha
}^{2}\right] }{r}_{+}^{{(3-n-{\alpha }^{2})/(\alpha }^{2}+1)},  \label{ddMS}
\end{equation}%
and%
\begin{equation}
\mathbf{H}_{S\mathbf{J}}^{M}=\frac{16\left( 1-\,{\alpha }^{2}\right) {l}%
^{2(n-4)}{r}_{+}^{2{(2-\,n)/(\alpha }^{2}+1)}}{{b}^{{\left( n-2\right)
\gamma }}{\Xi }^{4}\left[ \left( {\Xi }^{2}-1\right) {\alpha }^{2}+1+\left(
n-3\right) {\Xi }^{2}\right] }.  \label{HMQJ}
\end{equation}%
Since $\Xi ^{2}\geq 1$, both Eq. (\ref{ddMS}) and $\mathbf{H}_{S\mathbf{J}%
}^{M}$ are positive for $\alpha \leq 1$, therefore the uncharged rotating
solutions are stable in both canonical and grand-canonical ensembles
provided $\alpha \leq 1$. For this case, the temperature is also always
positive as one can see from (\ref{Texp}). As pointed out before, the charge
cannot change thermal stability and therefore we always have thermally
stable rotating black brane solutions for $\alpha \leq 1$. This fact is
illustrated in Figs. (\ref{fig3}) and (\ref{fig4}) for different values of $%
r_{+}$. The positivity of temperature for them is shown in Fig. (\ref{fig5}%
). For different $q$'s, Figs. (\ref{fig6}) and (\ref{fig7}) show that charge
does not affect the thermal stability and therefore charged solutions are
still stable for $\alpha \leq 1$. The positivity of $T$ for mentioned
parameters in Figs. (\ref{fig6}) and (\ref{fig7}) is shown in Fig. (\ref%
{fig8}).

Now, we discuss the stability for nonlinearly charged rotating black brane
solutions for $\alpha >1$. One can see from (\ref{HMQJ}) that $\alpha =1$ is
the root of $\mathbf{H}_{S\mathbf{J}}^{M}$. Numerical investigations show
that $\alpha =1$ is the root of determinant of Hessian matrix in charged
case too. Also, for $\alpha >1$, $\mathbf{H}_{SQ\mathbf{J}}^{M}$ is always
negative as $\mathbf{H}_{S\mathbf{J}}^{M}$ obviously is (see (\ref{HMQJ})).
Therefore, we have unstable solutions for $\alpha >1$ in grand-canonical
ensemble. Figs. (\ref{fig4}) and (\ref{fig7}) illustrate this fact. However,
in canonical ensemble we have both stable and unstable solutions for $\alpha
>1$. Figures (\ref{fig3}) and (\ref{fig6}) show that there is an $\alpha
_{\max }(>1)$ that we have stable solutions for values lower than it (note
that $\alpha <\sqrt{n-1}$; see sentences above (\ref{Mass})). There is also
a $r_{+\max }$ that for $r_{+}>r_{+\max }$ solutions are unstable (see Fig. (%
\ref{fig9})). The behavior of $(\partial ^{2}M/\partial S^{2})_{Q,\mathbf{J}%
} $ in terms of $q$ and $\beta $ are depicted in Figs. (\ref{fig10}) and (%
\ref{fig11}) respectively. These figures show that there are $q_{\min }$ and
$\beta _{\min }$ that for values greater than them black branes are
thermally stable.

\section{Conclusions and discussions}

In this paper, we studied the higher dimensional action in the context of
dilaton gravity and in the presence of the logarithmic nonlinear
electrodynamics. By varying the action, we found the field equations of this
theory. Then, we constructed a new class of charged, rotating black brane
solutions, with $k=[(n-1)/2]$ rotation parameters, in an arbitrary
dimension. We found that the presence of the dilaton field changes the
asymptotic behavior of the obtained solutions to be neither flat nor (A)dS.
We presented the suitable counterterm which remove the divergences of the
action in the presence of the dilaton field. In the absence of a non-trivial
dilaton ($\alpha =\gamma =0$), these solutions reduce to the asymptotically
AdS charged rotating black brane solutions of Einstein theory in the
presence of logarithmic nonlinear electrodynamics \cite{Hendi}. When $\beta
\longrightarrow \infty $, these solutions reduce to the charged rotating
dilaton black brane solutions given in Ref. \cite{SDRP}. We also calculated
the conserved and thermodynamic quantities of the spacetime such as mass,
angular momentum, temperature, entropy and electric potential and checked
that the first law of thermodynamics holds on the black brane horizon.

Then, we explored thermal stability of the nonlinearly charged
rotating black brane solutions in both canonical and
grand-canonical ensembles. We found that in both ensembles the
solutions are thermally stable for $\alpha \leq 1$, while for
$\alpha >1$ the solutions are always thermally unstable in the
grand-canonical ensemble {where }$\alpha $ {is the
dilaton-electromagnetic coupling constant.} In the canonical
ensemble, however, we can have both stable and unstable solutions
for $\alpha >1$. We found that, in this ensemble, there is an
$\alpha _{\max }>1$ for which the
solutions are thermally stable provided $\alpha <\alpha _{\max }$. {%
The pointed out results implies that the thermal stability is
ensemble-dependent and }$\alpha $ {influences the stability under
thermal perturbations. These results are expectable since
different ensembles allow different sets of quantities to be
variable and a thermally stable system is one which is stable
under varying variable quantities. On the other hand, values of
conserved and thermodynamic quantities depend on values of
parameters such as }$\alpha ${\ and therefore the fact that
dilaton-electromagnetic coupling has direct effect on thermal
stability of the system seems natural}.

It is notable to mention that in this paper, we only constructed the charged
rotating dilaton black branes of nonlinear electrodynamics with flat
horizon. One can try to construct the rotating dilaton black holes of this
theory with curved horizon. One can laso investigate the thermodynamic
geometry of these solutions. The latter issue is now under investigation and
the results will be presented elsewhere.

\textbf{Conflict of Interests}

The author declares no conflict of interests for the present paper.
\acknowledgments{We thank the Research Council of Shiraz
University. This work has been supported financially by Research
Institute for Astronomy \& Astrophysics of Maragha (RIAAM), Iran.}

\end{document}